\documentclass[pra,aps,twocolumn,superscriptaddress,showpacs]{revtex4}
\usepackage{graphicx}
\usepackage{amsmath}

\newcommand{\eq}{\begin{equation}}
\newcommand{\fine}{\end{equation}}

\begin{document}

\title{
\vspace{2cm
\begin{flushright}
CERN-PH-TH/2013-113
\end{flushright}
\vspace{3cm}
\bf \LARGE
Froissart Bound on Total Cross-section without Unknown Constants   }
\vspace{.2cm}}
\date{}
\author{Andr\'e Martin }
\email{martina@mail.cern.ch} \affiliation{Theoretical Physics Division,CERN, Geneva}

\author{S. M. Roy}
\email{smroy@hbcse.tifr.res.in} \affiliation{HBCSE,Tata Institute of Fundamental Research, Mumbai}

\begin{abstract}
We determine the scale of the logarithm in the Froissart bound on total cross-sections using 
absolute bounds on the D-wave below threshold for pion-pion scattering. 
E.g. for $\pi^0 \pi^0$ scattering we show that for c.m. energy $\sqrt{s}\rightarrow \infty $,  
$\bar{\sigma}_{tot }(s,\infty)\equiv s\int _{s} ^{\infty } ds'\sigma_{tot }(s')/s'^2   \leq \pi (m_{\pi })^{-2} 
[\ln (s/s_0)+(1/2)\ln \ln (s/s_0) +1]^2$ where $1/s_0= 17\pi \sqrt{\pi/2 }\>m_{\pi }^{-2}  $ .
   
\end{abstract}

\pacs{03.67.-a, 03.65.Ud, 42.50.-p}

\maketitle

{\bf Introduction}.
Froissart \cite{Froissart1961} proved from the Mandelstam representation that the total cross-section $\sigma_{tot} (s)$ for two particles to go to anything 
at c.m. energy $\sqrt s$ must obey the bound,
\begin{equation}
\sigma_{tot} (s) \leq_{s\rightarrow \infty} C \> [\ln (s/s_0)]^2 ,
\end{equation}
where $C, s_0$ are unknown constants.
Later Martin \cite{Martin1966} proved this bound rigorously from axiomatic field theory
by enlarging the Lehmann ellipse of analyticity \cite {Lehmann1958} for the absorptive part; further, the 
constant $C$ was fixed by Lukaszuk and Martin \cite{Lukaszuk-Martin1967} using unitarity and validity of 
dispersion relations with a finite number of subtractions for $-T< t\leq 0$ (and as a consequence, of,
twice subtracted fixed-$t$ dispersion relations for $|t|< t_0 $ \cite{Jin-Martin1964}), to obtain,
\begin{equation}
 \sigma_{tot} (s) \leq_{s\rightarrow \infty} 4\pi/(t_0 - \epsilon ) \> [\ln (s/s_0)]^2 
\equiv \sigma_{max} (s),
\end{equation}
where $t_0$ is the lowest singularity 
in the $t$-channel and $\epsilon $ an arbitrarily small positive constant. For many processes, for example for 
 $\pi \pi ,KK, K\overline K,\pi K, \pi N, \pi \Lambda$ scattering it is known \cite{Bessis-Glaser1967} that
$t_0 = 4 m_{\pi}^2 $,,  $m_{\pi} $ being the pion-mass . (We shall choose units $m_{\pi}=1$). 

These results were obtained by Martin \cite{Martin1966} in the framework of local field theory as applied to hadrons, 
using implicitly the Wightman axioms \cite{Wightman}. However, later, the needed analyticity properties , 
as well as polynomial boundedness at fixed momentum transfer, were obtained by Epstein, Glaser and Martin 
\cite{Epstein} in the more general framework of the theory of local observables of Haag, Kastler and Ruelle 
\cite{Haag}.

Recently, Azimov  has revisited the Froissart bound in a paper \cite{Azimov}, Sec. 2 of which is 
similar to the 1962 and 1963 works of Martin \cite{Martin1962}.These papers were a precursor to Martin's later 
paper \cite{Martin1966} which proved the bound rigorously from axiomatic field theory. Azimov has 
raised doubts about ``application of the ideas and methods of axiomatic local field theory 
to hadron properties''. His main point is that, ``hadrons, consisting of 
quarks and gluons, cannot be pointlike'', and might not be associated to local fields.
However, Zimmermann \cite{Zimmermann} has shown that local fields 
can be associated to composite particles (for instance deuterons). We postulate that this construction 
applies to hadrons made of quarks. This is not obvious because, in spite of the  practical successes
of QCD, nobody knows how to incorporate particles without asymptotic
fields in a field theory. Anyway this is a much weaker assumption than that of the validity of 
Mandelstam representation. In particular, we do not use the Froissart-Gribov representation of physical 
region partial waves for fixed s. 

The Froissart-Martin bound has triggered much work on high 
energy theorems (see e.g. \cite{Singh-Roy1970}, \cite{Roy1972}) and on models of high energy 
scattering \cite{Cheng-Wu1970}. Recently, Martin proved a  
bound on the total inelastic cross section at high energy \cite{Martin2009} which is 
one-fourth of the bound $\sigma_{max} (s) $ on the total cross-section.
Wu, Martin, Roy and Singh \cite{WMRS} obtained a bound on $\sigma_ {inel} (s) $ in terms of 
$\sigma_ {tot} (s)$ which vanishes both when the total cross-section vanishes and 
when it equals the unitarity upper bound.
  
In spite of all this progress, these bounds share severe shortcomings \cite{WMRS}.  
(i)They are deduced assuming that the absorptive part $A(s,t), 0\leq t< t_0 $ is bounded by Const.$ s^2 /\ln (s/s_0) $
for $s\rightarrow \infty $. In fact, the Jin-Martin theorem on twice subtracted dispersion relations 
only guarantees that 
\begin{equation}
 C(t) \equiv \int _{s_{th }} ^{\infty} ds A(s,t) /s^3 < \infty,\>0\leq t< t_0, 
\end{equation}
where $s_{th }$ is the $s$-channel threshold. As stressed by Yndurain and Common \cite{Yndurain}, this does not 
imply that $A(s,t) \leq Const. s^2/\ln (s/s_0) $ for all sequences of $s\rightarrow \infty $.
(ii)The bounds are expressed in terms of $\sigma_{max} (s) $ which still contains the unknown scale $s_0$ of 
the logarithm, and the unknown positive parameter $\epsilon $ which can be chosen arbitrarily small but $ \neq 0$. If 
$\epsilon $ is not fixed $s_0 $ cannot be fixed since the advantage of a larger $s_0 $ can be offset by 
a larger $\epsilon $.

We now remove both these shortcomings. We report definitive bounds on energy averages of the total
 cross-section in which the scale $s_0 $ is determined in terms of 
$C(t)$ which is a low energy (in fact below threshold) property in the $t-$channel. 
In some cases , e.g. for pion-pion scattering, for $ t \rightarrow 4$, $C(t)$ is proportional to the 
D-wave scattering length \cite{Colangelo2000} which is known phenomenologically; hence we obtain bounds 
on energy averages in terms of that scattering length. Even more exciting is the fact that for 
$\pi^0 \pi^0$ scattering we are able to obtain absolute bounds (in terms of pion-mass alone) on $C(t)$ below 
threshold without assuming finiteness of the D-wave scattering length; this yields absolute bounds on the 
asymptotic energy averages of the total cross-section. 
     
{\bf Normalizations}.
Let $F(s,t)$ be an $ab\rightarrow ab$ scattering amplitude  at c.m. energy $\sqrt{s}$ 
and momentum transfer squared $t$  normalized for non-identical partcles $a,b$ such that 
the differential cross-section is given by
\begin{equation}
 \frac{d \sigma }{d \Omega } (s,t) = \bigl |4 \frac{F(s,t)} {\sqrt{s} } \bigr |^2 ,
\end{equation}
 with $t$ being given in terms of the c.m. momentum $k$ and the scattering angle $\theta$ by the 
relation,
\begin{equation}
 t= -2 k^2 (1- \cos \theta ); \>  z \equiv \cos \theta= 1+ t / (2 k^2).
\end{equation}
Then, for fixed $s$ larger than the physical $s-$channel threshold, $F(s;\cos \theta ) \equiv F(s,t)$ is 
analytic in the complex $\cos \theta $ -plane inside the Lehmann-Martin ellipse with foci -1 and +1 and 
semi-major axis $\cos \theta _0 = 1+ t_0 /(2 k^2) $.
Within the ellipse ,in particular, for $ |t| < t_0 , F(s,t)$ and the $s$-channel absorptive 
part $F_s(s,t)=A(s,t) $ have 
the convergent partial wave expansions,
\begin{equation}
  F(s,t)=\frac{ \sqrt s }{4k} \sum_{l=0}^{\infty} (2l+1)P_l (z)a_l (s),
\end{equation}
\begin{equation}
 F_s(s,t)=A(s,t)=\frac{ \sqrt s }{4k} \sum_{l=0}^{\infty} (2l+1)P_l (z) Im a_l (s),
\end{equation}
with the unitarity constraint,
\begin{equation}
 Im a_l (s) \geq | a_l (s) |^2 ,\> s\geq 4 \>.
\end{equation}
Correspondingly, the optical theorem gives, for $a\neq b$,
\begin{eqnarray}
 \sigma _{tot} (s) = \frac{4 \pi}{k}Im \big (4F(s,0)/\sqrt s \big )\nonumber\\
 =\frac{4 \pi}{k^2} \sum _{l=0}^{\infty} (2l+1) Im a_l(s)\>.
\end{eqnarray}
For identical particles $a=b$ e.g. for $\pi^0 \pi^0$ scattering, or for pion-pion scattering with Iso-spin $I$, 
we have the same formula for the differential cross-section,
$$\frac{d \sigma }{d \Omega } (s,t) = \bigl |4 \frac{F(s,t)} {\sqrt{s} } \bigr |^2 ,$$ and the same 
form of the unitarity constraint,
\begin{equation}
 Im a_l^I (s) \geq | a_l^I (s) |^2 ,\> s\geq 4 \>,
\end{equation}
but the partial waves $ a_l (s) \rightarrow 2 a_l^I (s) $ in the partial wave expansion,i.e.
\begin{equation}
 F^I(s,t)=\frac{ \sqrt s }{4k} \sum_{l=0}^{\infty} (2l+1) 2 a_l^I (s)P_l (z).
\end{equation}
With this normalization, $F^I (4,0)= a_0^I $, the S-wave scattering length for Iso-spin $I$. 
and for pion-pion scattering the identical particle factors lead to,
\begin{equation}
 \sigma _{tot}^I (s) =  \frac{4 \pi}{k^2} \sum _{l=0}^{\infty} (2l+1) 2 Im a_l^I(s) \>.
\end{equation}
In the following, we shall consider non-identical particles $a\neq b$ for detailed derivations and quote 
the identical particle results when needed.

{\bf Convexity Properties of Lower Bound on Absorptive Part in terms of Total Cross-Section}. 
Martin has proved 
unitarity lower bounds on $A(s,t) $ for $ 0<t<t_0$ in terms of $\sigma _{tot} (s)$
\cite{Martin1966},and in terms of $\sigma _{inel} (s)$\cite{Martin2009}. He has also proved \cite{Martin2010} 
that these 
bounds are convex functions of $\sigma _{tot} (s)$, and $\sigma _{inel} (s)$ respectively.
We recall first the convexity properties which will be crucial for our proofs of lower bounds on $C(t)$ 
in terms of energy averages of total cross-sections.
We work at a fixed-$s$ , and suppress the $s$-dependence of $Im a_l (s)$,
 and $ \sigma _{tot} (s) $ for simplicity of writing.
Using $0\leq Im a_l \leq 1 $, the lower bound on $A(s,t) $ for given $ \sigma _{tot} $ is obtained by choosing,
\begin{eqnarray}
 Im a_l = 1, \> 0 \leq l \leq L \>;  Im a_{L+1 } = \eta \>; \nonumber\\
Im a_l =0,  \> l > L+1 ,
\end{eqnarray}
where, the fraction $\eta , 0\leq \eta <1$, and the integer $L $ are determined from the given $ \sigma _{tot}$. Thus,
\begin{eqnarray}
A(s,t)\frac{4k}{\sqrt s }\geq &\big (\sum_{l=0}^{L} (2l+1) P_l (z) + \eta (2L+3) P_{L+1 }(z)\big )\nonumber\\
     \equiv &A (z) ,
\end{eqnarray}
 where,
\begin{equation}
 \sigma _{tot} \frac{ k^2}{4\pi}   = \big (\sum_{ l=0}^{l=L } (2l+1) +\eta (2L+3)\big )\equiv \Sigma_{tot }.
\end{equation}
Hence, $A(z)$ is a monotonically increasing function of $\Sigma_{tot } $ with piecewise constant  positive 
derivative. Denoting $Int (x)= $ integer part of $x$,
\begin{equation}
 dA(z)/d\Sigma_{tot }= P_{L+1 }(z), \> L= Int (\sqrt{\Sigma_{tot } })-1 ,
\end{equation}
which increases with $L$ since $z>1$, and hence with $\Sigma_{tot }$ when it crosses square of an integer.
This proves that the lower bound $A(z)$ is a convex function of $\Sigma_{tot }$, and that,
\begin{equation}
 A(z)=\Sigma_{tot }, \>for\>\Sigma_{tot } \leq 1,
\end{equation}
and, for $ \Sigma_{tot } > 1 $
\begin{eqnarray}
A(z) = 1+ \int _1 ^{\Sigma_{tot} }P_{Int (\sqrt{\sigma} ) } (z) d \sigma , \> \nonumber\\
\geq 1+2\int _0 ^{ \sqrt{\Sigma_{tot } }-1 } (\mu +1) P_{\mu } (z) d\mu .
\end{eqnarray}
Using integral representations for $P_{\mu } (z)$ and for the modified Bessel function  $ I_0$ we obtain 
for  $\mu \geq 0, \> z>1$,
\begin{equation}
 P_{\mu } (z) \geq  I_0 (\mu \ln z_+), \> z_+ \equiv z+ \sqrt{z^2-1}. 
\end{equation}
This yields the strict inequality ( without any high energy approximation ),
\begin{eqnarray}
 A(z)\geq  2 \big (\frac{xI_1(x) } {(\ln z_+)^2 } + \frac{I_0 (x) } {\ln z_+ } \big )|_ 
{x= (\sqrt{\Sigma_{tot }} -1)\ln z_+} +\nonumber\\
1+2 \ln z_+ , \>for \> \Sigma_{tot } > 1 \>.
\end{eqnarray}
At high energy, this gives,
\begin{equation}
A(s,t) >\frac{s } {4t} xI_1(x)\big |_{x=\sqrt{t\sigma_{tot }(s)/(4 \pi) } } (1 + O(1/\sqrt{s }))
\end{equation}
which is a convex function of $\sigma_{tot }(s) $.

{\bf Upper bound on energy-averaged total cross-section }. Defining,
\begin{equation}
 \bar{\sigma}_{tot }(s,\infty)\equiv s\int _{s} ^{\infty } \frac{ds' } {s'^2 } \sigma_{tot }(s'),
\end{equation}
and 
\begin{equation}
 C_s(t)  \equiv \int _{s} ^{\infty} ds' A(s',t) /s'^3 < \infty,\>0\leq t< t_0,
\end{equation}
we obtain,
\begin{eqnarray}
 C_s(t) \geq \frac{1 } {4ts } s\int _{s} ^{\infty } \frac{ds' } {s'^2 }\sqrt{\frac{t\sigma_{tot }(s')}{4 \pi} } 
I_1(\sqrt{\frac{t\sigma_{tot }(s')}{4 \pi} } )\nonumber\\
\geq \frac{1 } {4ts }\sqrt{\frac{t\bar{\sigma}_{tot }(s,\infty)}{4 \pi} } 
I_1(\sqrt{\frac{t\bar{\sigma}_{tot }(s,\infty)}{4 \pi} } ),
\end{eqnarray}
since the average of a convex function must be greater than the convex function of the 
average \cite{Hardy}. 
At high energies if $\bar{\sigma}_{tot }(s,\infty) $ goes to $\infty$, the asymptotic expansion of $I_1 (\xi ) $ 
yields,
\begin{equation}
 4st C_s(t) \sqrt{2 \pi } > \big (\sqrt{\xi }\exp{\xi }\big )(1 + O(1/\xi)), 
\end{equation}
where,
\begin{equation}
 \xi \equiv \sqrt{\frac{t\bar{\sigma}_{tot }(s,\infty)}{4 \pi} }.
\end{equation}
To extract a bound on the cross-section, we need the following lemma \cite{Martin2010}.
If $\xi >1 $, and
\begin{equation}
  y \geq \sqrt{\xi }\exp{\xi }, 
\end{equation}
then, 
\begin{equation}
 \xi< f(y)\equiv \ln{ y} -(1/2)\ln {\big (\ln{ y}-\frac{1}{2}\ln {\ln{ y} } \big )}.
\end{equation}
Proof. It is enough to prove this for $y = \sqrt{\xi }\exp{\xi }$, since the right-hand side 
is an increasing function of $\xi $.Taking logarithms ,and using $\xi =\ln{ y} -(1/2)\ln {\xi }\equiv \xi _1$
 repeatedly,
\begin{equation}
 \xi =\ln{ y} -(1/2)\ln {\big (\ln{ y}-\frac{1}{2}\ln {\xi _1 } \big )}.
\end{equation}
For fixed $y$ the derivative of the right-hand side with respect to $\xi _1$ is $(4 \xi _1^2)^{-1}$ which is 
positive, and $\xi _1 < \ln {y} $ for $\xi >1 $. Hence the stated upper bound on $\xi $ follows.

Instead of the s-dependent $C_s(t)$ we shall use the simple $s$-independent upper bound ,
\begin{equation}
 C_s(t)\leq  C(t)- \int _{4} ^x ds' \frac {k'\sqrt{s'} \sigma_{tot }(s')}{(s')^3 16 \pi} ,\>4 < x< s
\end{equation}
which follows by using $A(s,t)> A(s,0) $ for $4>t>0$ and improves the value $C(t)$ if low energy
total cross-sections are known. The integral of the weight function multiplying $\sigma_{tot } $ can be done. Thus,
\begin{equation}
 C_s(t)\leq C_x(t) \equiv C(t)-\frac{(x-4)^{3/2}\bar{\sigma}_{tot }(x) } {12 x^{3/2}16\pi },
\end{equation}
where,
\begin{equation}
 \bar{\sigma}_{tot }(x) =\frac{\int _{4} ^x ds' k'\sqrt{s'} \sigma_{tot }(s')/s'^3 }
{\int _{4} ^x ds' k'\sqrt{s'} /s'^3 }
\end{equation}

With $f(y)$ as defined above, the upper bound on the average total cross-section in terms of $C_x(t) $is,
\begin{eqnarray}
 \bar{\sigma}_{tot }(s,\infty) \leq _{s\rightarrow \infty }\frac{4\pi } {t } \big ( f(s/s_0) +O(\ln (s/s_0))^{-1}
\big )^2\> ,\nonumber\\
 \frac{1 } { s_0}=4t C_x(t) \sqrt{2\pi },\> t=4m_\pi ^2 -\epsilon .
\end{eqnarray}

We may also find bounds on the average of the total cross-section in the interval $(s,2s) $,
\begin{equation}
 \bar{\sigma}_{tot }(s,2s)\equiv 2s\int _{s} ^{2s } \frac{ds' } {s'^2 } \sigma_{tot }(s').
\end{equation}
The lower bound on $A(s,t) $ and its convexity yield,
\begin{eqnarray}
 C_x(t) \geq \frac{1 } {8ts } 2s\int _{s} ^{\infty } \frac{ds' } {s'^2 }\sqrt{\frac{t\sigma_{tot }(s')}{4 \pi} } 
I_1(\sqrt{\frac{t\sigma_{tot }(s')}{4 \pi} } )\nonumber\\
\geq \frac{1 } {8ts }\sqrt{\frac{t\bar{\sigma}_{tot }(s,2s)}{4 \pi} } 
I_1(\sqrt{\frac{t\bar{\sigma}_{tot }(s,2s)}{4 \pi} } ).
\end{eqnarray}
Asymptotically we obtain a bound of the same form as before, but with the scale factor in the logarithm being 
$s_0/2 $,
\begin{equation}
 \bar{\sigma}_{tot }(s,2s) \leq _{s\rightarrow \infty }\frac{4\pi } {t } \big ( f(2s/s_0) +O(\ln (s/s_0))^{-1}
\big )^2\>.
\end{equation}

Note that  $\sigma _{tot }(s) <
\bar{\sigma}_{tot }(s,2s)$ if the cross-section increases with $s$ in the interval $(s,2s) $;the above 
bound on energy averages therefore immediately 
yields a bound on $\sigma _{tot }(s)$ in that case.

For identical particles there are only even partial waves in the partial wave expansions, but the  
lower bound on the absorptive part is again a convex function of the total cross-section; the identical 
particle factors multiplying the partial waves ensure that inspite of only even partial waves contributing,
the largest partial wave $L$ in the variational bound which is of $O(\sqrt{s\sigma_{tot }(s) }) $ has only 
$O(1)$ corrections with respect to the non-identical particles case.The quoted asymptotic bounds on the absorptive 
part in terms of $\sigma_{tot } $ and on the energy averaged total cross-section in terms of $C_x(t) $  therefore 
remain unchanged. 

{\bf Phenomenological Bounds in terms of D-wave Scattering Length}. Rigorous results from axiomatic field theory 
do not guarantee finiteness of the D-wave scattering lengths. However if we use phenomenological values for them 
we can choose $\epsilon =0$ and evaluate $C(t=4)$ . We shall use, $F^{\pi^+ \pi^0\rightarrow \pi^+ \pi^0}=
1/2( F^1 + F^2) $, $F^{\pi^0 \pi^0\rightarrow \pi^0 \pi^0}= \frac{1}{3} F^0 + \frac{2}{3}F^2 $, 
the crossing relation, 
\begin{equation}
\frac{1}{2}( F^1 + F^2)(s,t)=\frac{1}{3}(F^0-F^2)(t,s),
\end{equation}
and the total crossing symmetry of the $\pi^0 \pi^0\rightarrow \pi^0 \pi^0 $ amplitude.
If we denote $F(s,t)=G(t,s)=G(t;z_t)$ where $F(s,t)$ denotes the  $\pi^+ \pi^0\rightarrow \pi^+ \pi^0 $ 
or the $\pi^0 \pi^0\rightarrow \pi^0 \pi^0$ amplitude , then the corresponding $G(t,s)$ has only even partial waves, 
\begin{equation}
 g_l(t)= \frac{1} {2 }\int _{-1 }^1 dz_t P_l (z_t) G(t;z_t).
\end{equation}
For $0<t<4 $, with the absorptive part $F_s (s,t)$ defined by Eq. (7), the fixed-$t$ dispersion relations 
with two subtractions imply the Froissart-Gribov formula rigorously for  $l\geq 2$,
\begin{equation}
  g_l(t)=\frac{4 } {\pi (4-t)}\int_4^{\infty }ds'Q_l\big (\frac{2(s'-4)+4+t } {4-t } \big)F_s (s',t),
\end{equation}
where $Q_l $ denotes the Legendre function of the second kind.The positivity of the absorptive part 
then implies the positivity of $g_l(t) $ for $ 0<t<4$; further, 
\begin{equation}
g_2(t)\rightarrow _{t\rightarrow 4 } \frac{(t-4)^2 } {15\pi }\int_4^{\infty }ds'\frac{F_s (s',t)}{s'^3}.
\end{equation}
If the $t$-channel $D$-wave scattering lengths exist,the definitions of $C(t)$ and of 
the $D$-wave scattering lengths $a_2^I $for iso-spin $I$ yield,
\begin{equation}
 C^{\pi^+ \pi^0\rightarrow \pi^+ \pi^0 }(t=4)= \frac{5\pi } {16 }m_\pi (a_2^0 -a_2^2),
\end{equation}
and 
\begin{equation}
 C^{\pi^0 \pi^0\rightarrow \pi^0 \pi^0 }(t=4)= \frac{5\pi } {16 }m_\pi (a_2^0 +2a_2^2).
\end{equation}
Here we have defined the $l$-wave scattering lengths $a_l^I$ as the $q \rightarrow 0$ limits of 
the phase shifts $\delta _l ^I (q) $ divided by $q^{2l+1}$ where $q$ is the c.m. momentum . 
Then an $S$-wave scattering length is indeed a length, with dimension $m_\pi ^{-1}$, and the 
$D$-wave scattering lengths have dimension $m_\pi ^{-5}$. Then, phenomenologically \cite{Colangelo2000} we have,
\begin{equation}
 a_2^0\approx 0.00175 m_\pi ^{-5 }\>;a_2^2\approx 0.00017 m_\pi ^{-5 }.
\end{equation}
and Roy \cite{Roy1972} has obtained from low energy data, for $x=50$,
\begin{equation}
 \bar{\sigma}_{tot }^{\pi^0 \pi^0 }(x)=8.2\pm 4\> mb ;\>\bar{\sigma}_{tot }^{\pi^+ \pi^0 }(x)=17\pm 3.5 \>mb.
\end{equation}

With $\epsilon =0$ ,$t=4$ and the values of $C_x(t=4)$ given in terms of the scattering lengths, and 
the low energy total cross-sections, we have, from Eqs. (31)-(33), with $x=50$,
\begin{eqnarray}
\pi^0 \pi^0&:&\>s_0=17\> m_\pi ^2 ,\nonumber\\
 C_x(4)&=&2.05\times 10^{-3}-0.6\times 10^{-3}\nonumber\\
&=&1.45\times 10^{-3} \>m_\pi ^{-4 }.\nonumber\\
\pi^+ \pi^0 &:&\>s_0=81\> m_\pi ^2 ,\nonumber\\
C_x(4)&=&1.55\times 10^{-3}-1.24\times 10^{-3}\nonumber\\
&=&.31\times 10^{-3} \>m_\pi ^{-4 } 
\end{eqnarray} 
where we have indicated the separate contributions of the D-wave scattering lengths and low energy total cross-sections 
to $C_x(4)$ but have not indicated the (substantial) errors on them which imply corresponding errors 
on the scale factors. Our bounds on 
average total cross-sections for $\pi^+ \pi^0 $ and $\pi^0 \pi^0 $ scattering therefore do not contain 
any unknown constants but the scale factor $s_0$ has large phenomenological errors. We cure this problem in the 
next section at the cost of getting poorer bounds.

{\bf Absolute bounds on the D-wave below threshold for $\pi^0 \pi^0$ 
scattering}. Although threshold behaviour cannot be proved from first principles, it was shown long ago 
\cite{Martin1967} that  $|f_l(s)|< C (4-s)^{l-1} $ must hold for $0<s<4$. We derive an absolute bound of this 
form and use it to derive a rigorous asymptotic bound on energy averaged  
total cross-section for $\pi^0 \pi^0$ scattering without unknown constants.As noted already, for $0<s<4 $ and
 $l\geq 2 $, the Froissart-Gribov formula implies that $f_l(s)>0 $. Hence, for $0<s<4,4-s<t<4 $ the convergent 
partial wave expansion, 
\begin{eqnarray}
F(s,t)-F(s,0)= \nonumber\\
\Sigma _{l=2}^{\infty } (2l+1)f_l(s) \big ( P_l(\frac{2t-4+s}{4-s}) -1 \big ) 
\end{eqnarray}
is in fact a sum of positive terms and yields an upper bound for the $l\geq 2 $ partial waves if we can 
obtain a bound on $F(s,t)-F(s,0) $ using analyticity.
The twice subtracted fixed-$t$ dispersion relations in s can be rewritten in terms of the convenient 
variable $z\equiv (s-2+t/2)^2 $, with $F(s,t)\equiv F(z;t) $. For $0\leq t <4$, 
\begin{equation}
F(s,t)-F(\frac{4-t } {2 },t)=\frac{z } {\pi } \int_{z_0 }^{\infty }dz'\frac{Im F(z';t)}{z'(z'-z)}\>,
\end{equation}
and the positivity of the absorptive part then yields,
\begin{equation}
 F(s,t)-F(\frac{4-t } {2 },t) \geq 0,if\>0\leq z <z_0 =(2+\frac{t}{2})^2 .
\end{equation}
If $s_1<s<4$ and $z_1\equiv (s_1-2+t/2)^2$, then
\begin{equation}
 z_1-z=(s_1-s)(s_1+s-4+t)< 0,if\>t > 4-s-s_1,
\end{equation}
and hence for $z_1<z<z_0 $,
\begin{eqnarray}
(z'-z)^{-1}-(z_0-z_1)\big ((z_0-z )(z'-z_1)\big )^{-1}=\nonumber\\
(z'-z_0 )(z_1-z )\big ((z'-z )(z_0-z )(z'-z_1)\big )^{-1} < 0 \>.
\end{eqnarray}
Inserting this into the dispersion relation we have,for $4>t > 4-s-s_1  $, and $t\geq 0 $,
\begin{eqnarray}
  F(s,t)-F(\frac{4-t } {2 },t)< \frac{(4-s_1)(s_1+t)^2 (s-2+\frac{t}{2})^2}{(4-s)(s+t)^2 (s_1-2+\frac{t}{2})^2}
\nonumber\\
\big (F(s_1,t)-F(\frac{4-t } {2 },t)\big ),if\>s_1<s<4\>.
\end{eqnarray}
Choosing $s_1=3, t=2$ and $3<s<4$, we get
\begin{equation}
 F(s,2)-F(1,2)<(25/16)(F(3,2)-F(1,2) )/(4-s).
\end{equation}
Using this and $F(s,0)>F(2,0) $
\begin{eqnarray}
F(s,2)-F(s,0)<F(1,2)-F(2,0) +\nonumber\\
(25/16)(F(3,2)-F(1,2) )/(4-s) ,for\>3<s<4 .
\end{eqnarray}
We now use absolute bounds on pion-pion amplitudes first discovered by Martin \cite{Martin1965}, and 
improved successively by \cite{Lukaszuk-Martin1967}, \cite{Auberson et al} and  
\cite{Lopez-Mennessier} in the improved final form,
\begin{eqnarray}
 -7.25<F(1,2)<2.75,\nonumber\\
F(2,0)>-3.5,F(3,2)<14.5
\end{eqnarray}
with normalization $F(4,0)=$ S-wave scattering length,and obtain the absolute bound,
\begin{equation}
 F(s,2)-F(s,0) < 6.25+\frac{33.99 } {4-s }\>,for \>3<s<4.
\end{equation}
The partial wave expansion of $F(s,2)-F(s,0) $ now yields for $3<s<4$,
\begin{equation}
f_l(s)\leq \frac{6.25+\frac{33.99 } {4-s } }{(2l+1) \big ( P_l(\frac{s}{4-s}) -1 \big ) }\>,
\end{equation}
which implies in particular,
\begin{equation}
f_2(s)<_{s\rightarrow 4- } \frac{4-s}{120}\big (34+6.25(4-s)+O(4-s)^2 \big ).
\end{equation}
With $s$ replaced by $t$ in this formula, the Froissart-Gribov formula now yields,
\begin{equation}
  C^{\pi^0 \pi^0\rightarrow \pi^0 \pi^0 }(t)<_{t\rightarrow 4- }\frac{17 \pi}{4(4-t)}
\end{equation}

{\bf Absolute bound on energy averaged total cross-section for $\pi^0 \pi^0$ 
scattering at high energy}. Inserting the bound on $C(t)$ into the average cross-section bound, the 
optimum value of $t$ turns out to be $t=4-(1/8 \ln (s/s_0))^{-1 } $, and the optimum bound,
\begin{eqnarray}
\bar{\sigma}_{tot }(s,\infty)  &\leq& \pi (m_{\pi })^{-2} 
[\ln (s/s_0)+(1/2)\ln \ln (s/s_0) +1]^2\nonumber\\
&+&O(\ln \ln (s/s_0)), \> \> s_0^{-1}=17\pi \sqrt{\pi/2 }\> m_{\pi }^{-2}. 
\end{eqnarray}
For $\bar{\sigma}_{tot }(s,2s)$ we obtain the same form of the bound ,but with half the value of $s_0$.

{\bf Outlook and Acknowledgements}.
 Our basic bound on the absorptive part, Eq. (20), is valid at all energies and its energy integral may be used for 
comparisons with experimental cross-section data which have a large non-asymptotic contribution 
at current energies. We have highlighted the simpler asymptotic upper bounds on average total cross-sections . 

We believe that our result 
is important as a matter of principle. However, we also believe that the
magnitude of the coefficient in front of the Froissart bound is not
satisfactory, especially if one decides to beiieve that the Froissart term is universal and compares with
p-p and p-pbar cross-sections at the ISR \cite{ISR},at the SppbarS \cite{SppbarS}, at the Tevatron
\cite{Tevatron} and at the LHC \cite{LHC}. All these
indicate the existence of a Froissart like contribution with a much smaller coefficient, and a much larger
scale and are well reproduced by, for instance, the BSW model \cite{Cheng-Wu1970} which incorporates automatically the
Froissart behaviour. Returning to $\pi\pi$ scattering, can the situation be improved? Yes, because one has to enforce
crossing symmetry and unitarity. Kupsch \cite{Kupsch} has constructed a crossing symmetric model satisfying Eq.(8), but
never tried to get numerical results. Also, we believe that unitarity in the elastic strips could be important.
This led to the discovery by Gribov \cite{Gribov} that the behaviour sF(t) for the total amplitude is impossible. 
If you remove the elastic unitarity constraint \cite{Martin-Richard} the
Gribov theorem disappears. To attack the problem one could use a variational approach taking as
an input the inelastic double spectral function in the Mandelstam representation. All we need is to find someone
courageous not looking for a job.

Similar bounds on inelastic cross-sections without any unknown constants will be reported separately 
\cite{Martin-Roy}. 

{\bf Acknowledgements}. One of us (A.M.) would like to thank the members of ITEP (Moscow) for 
inviting him to give a talk in 2010 which resurrected his interest in the 
subject, and also Tai Tsun Wu and Raymond Stora for stimulating discussions; the other (SMR) wishes to thank the Indian National Science Academy for the 
INSA Senior Scientist award, and L. Alvarez-Gaume for an invitation to CERN in 2009 which helped start 
this collaboration.

\end{document}